\documentclass[runningheads]{llncs}
\usepackage{amsmath}
\usepackage{graphicx}
\usepackage{tabularx}
\usepackage{booktabs} 
\usepackage{hyperref}
\usepackage{amssymb}
\usepackage{algorithm}
\usepackage{algpseudocode}
\usepackage{blindtext}

\DeclareMathOperator*{\argmax}{arg\,max}
\makeatletter
\def\blfootnote{\xdef\@thefnmark{}\@footnotetext}
\makeatother

\begin{document}
\title{BiX-NAS: Searching Efficient Bi-directional Architecture for Medical Image Segmentation}
\titlerunning{BiX-NAS for Medical Image Segmentation}
%

\author{Xinyi Wang\inst{1,*} \and
Tiange Xiang\inst{1,*} \and
Chaoyi Zhang\inst{1} \and
Yang Song\inst{2} \and
Dongnan Liu \inst{1} \and
Heng Huang \inst{3,4} \and
Weidong Cai \inst{1}
}

\authorrunning{Wang et al.}
%
\institute{School of Computer Science, University of Sydney, Australia \and
School of Computer Science and Engineering, University of New South Wales, Australia \and
Electrical and Computer Engineering, University of Pittsburgh, USA \and 
JD Finance America Corporation, Mountain View, CA, USA \\
\email{\{xwan2191, txia7609\}@uni.sydney.edu.au} \\
\email{\{chaoyi.zhang, tom.cai, dongnan.liu\}@sydney.edu.au} \\
\email{yang.song1@unsw.edu.au}\\
\email{henghuanghh@gmail.com}}

\maketitle              
\begin{abstract}
The recurrent mechanism has recently been introduced into U-Net in various medical image segmentation tasks. Existing studies have focused on promoting network recursion via reusing building blocks. Although network parameters could be greatly saved, computational costs still increase inevitably in accordance with the pre-set iteration time. In this work, we study a multi-scale upgrade of a bi-directional skip connected network and then automatically discover an efficient architecture by a novel two-phase Neural Architecture Search (NAS) algorithm, namely BiX-NAS. Our proposed method reduces the network computational cost by sifting out ineffective multi-scale features at different levels and iterations. We evaluate BiX-NAS on two segmentation tasks using three different medical image datasets, and the experimental results show that our BiX-NAS searched architecture achieves the state-of-the-art performance with significantly lower computational cost. Our project page is available at: \url{https://bionets.github.io}.

\end{abstract}
\blfootnote{* Equal contributions.}
\section{Introduction}
Deep learning based methods have prevailed in medical image analysis. U-Net \cite{ronneberger2015u}, a widely used segmentation network, constructs forward skip connections (skips) to aggregate encoded features in encoders with the decoded ones. Recent progress has been made on the iterative inference of such architecture by exploiting the reusable building blocks.
\cite{wang2019recurrent} proposed a recurrent U-Net that recurses a subset of paired encoding and decoding blocks at each iteration.
BiO-Net \cite{xiang2020bio} introduces backward skips passing semantics in decoder to the encoder at the same level. Although this recurrent design could greatly slim the network size, the computational costs still increase inevitably in accordance with its pre-set iteration time. Meanwhile, the success of multi-scale approaches \cite{zhou2018unetpp,huang2020unet3+} suggests the usage of multi-scale skips that fuse both fine-grained traits and coarse-grained semantics. 
To this end, introduction of forward/backward skips from multiple semantic scales and searching efficient aggregation of multi-scale features with low computational cost is of high interest.

Neural Architecture Search (NAS) methods automatically perceive the optimal architecture towards effective and economic performance gain. Classic evolutionary NAS methods \cite{real2017large,real2019regularized} evolve by randomly drawing samples during searching and validating each sampled model individually. Differentiable NAS algorithms \cite{liu2018darts,guo2020single} relax the discrete search space to be continuous and delegate backpropagation to search for the best candidate. Auto-DeepLab \cite{liu2019auto} applied differentiable NAS into image segmentation tasks to determine the best operators and topologies for each building block. Concurrently, NAS-Unet \cite{weng2019unet} applied an automatic gradient-based search of cell structures to construct an U-Net like architecture.
Despite the success on feature fusions at same levels, \cite{yan2020ms} introduced a multi-scale search space to endow their proposed MS-NAS with the capability of arranging multi-level feature aggregations. However, architectures searched by the above NAS algorithms only bring marginal improvements over hand-designed ones and their searching process is empirically inefficient. 

In this paper, we present an efficient \textbf{multi-scale} (abstracted as 'X') NAS method, namely BiX-NAS, which searches for the optimal \textbf{bi-directional} architecture (BiX-Net) by recurrently skipping multi-scale features while discarding insignificant ones at the same time.

Our contributions are three-fold: 
\textbf{(1)} We study the multi-scale fusion scheme in a recurrent bi-directional manner, and present an effective two-phase Neural Architecture Search (NAS) strategy, namely BiX-NAS, that automatically searches for the optimal bi-directional architecture. 
\textbf{(2)} We analyze the bottleneck of the searching deficiency in classic evolution-based search algorithms, and propose a novel progressive evolution algorithm to further discover a subset of searched candidate skips and accelerate the searching process.
\textbf{(3)} Our method is benchmarked on three medical image segmentation datasets, and surpasses other state-of-the-art counterparts by a considerable margin.

\noindent
\section{Methods}
We first discuss the effectiveness of introducing multi-scale skip connections to BiO-Net as an intuitive upgrade (Sec \ref{bionet++}), then we demonstrate the details of each phase of BiX-NAS (Sec \ref{exnas}), and eventually, we present the \textit{skip fairness} principle which ensures the search fairness and efficiency (Sec \ref{analysis}). 

\begin{figure}[t]
	\begin{center}
		\includegraphics[width=0.9\linewidth]{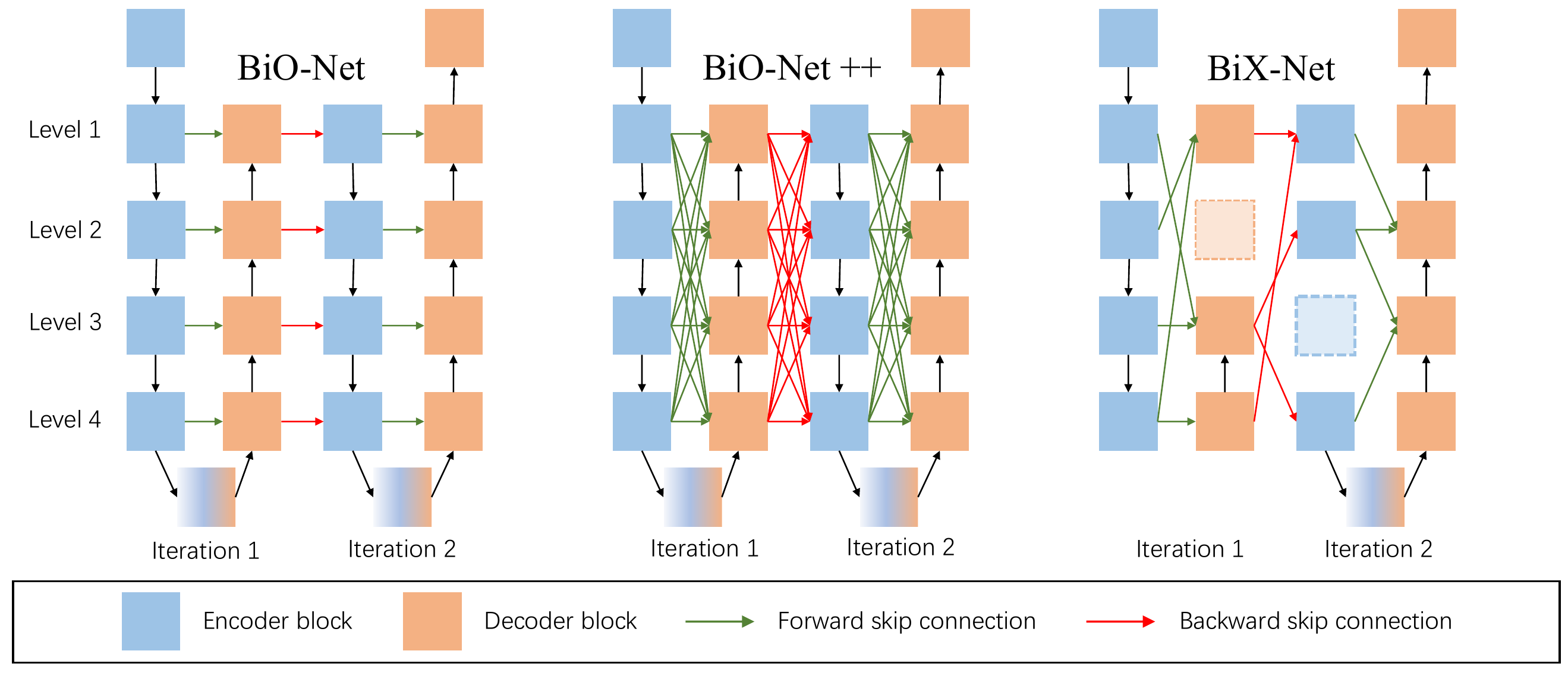}
	\end{center}
	\caption{\textbf{Overview of BiO-Net, BiO-Net++, and BiX-Net} with 4 levels and 2 iterations. Encoder and decoder blocks at the same level are reused \cite{xiang2020bio}.}
	\label{fig:1}
\end{figure}

\noindent
\subsection{BiO-Net++: A Multi-scale Upgrade of BiO-Net} \label{bionet++}
BiO-Net \cite{xiang2020bio} triggers multiple encoding and decoding phases that concatenate skipped features at the same level (Fig. \ref{fig:1} left). We optimize the feature fusion scheme as element-wise average for an initial reduction on the total complexity. For better clarity, we denote (1) A sequential encoding or decoding process as an \textit{extraction stage}. There are four extraction stages in a BiO-Net like network with two iterations. (2) Any blocks in a non-first extraction stage as \textit{searching blocks}. To fuse multi-scale features, precedent encoded/decoded features at all levels are densely connected to every decedent decoding/encoding level through bi-directional skips. We align inconsistent spatial dimensions across different levels via bilinear resizing. The suggested BiO-Net++ is outlined in Fig. \ref{fig:1} middle.

Although the above design promotes multi-scale feature fusion and shrinks network size, empirically we found that such \textit{dense} connections bring a mere marginal improvement in terms of the overall performance but with an increase of computational costs (Table \ref{fig:1}).
We are interested in seeking a \textit{sparser} connected sub-architectures of BiO-Net++ which could not only benefit from multi-scale fusions but also ease computation burdens to the greatest extent.

\subsection{BiX-NAS: Hierarchical Search for Efficient BiO-Net++} \label{exnas}
To this end, we present BiX-NAS, a two-phase search algorithm, to find a sparsely connected sub-architecture of BiO-Net++, where a trainable selection matrix is adopted to narrow down the search space for differentiable NAS in Phase1, and evolutionary NAS is introduced to progressively discover the optimal sub-architecture in Phase2.
To spot an adequate candidate, dense skip connections between every pair of extraction stages are further sifted for better efficiency. Suppose there are $N$ incoming feature streams in a desired searching block ($N=5$ for BiO-Net++), we anticipate that in a sparser connected architecture, only $k \in [1, N-2]$ candidate(s) of them could be accepted to such blocks, which results in a search space of $\approx\sum_{k=1}^{(N-2)}\binom{N}{k}^{\mathbf{L}(2\mathbf{T}-1)}$ in the SuperNet BiO-Net++ with $\mathbf{L}$ levels and $\mathbf{T}$ iterations. When $\mathbf{L}=4$ and $\mathbf{T}=3$ \cite{xiang2020bio}, the search space expands to $5^{40}$, escalating the difficulty to find the optimal sub-architectures.

\noindent
\textbf{Phase1: Narrowing down search space via selection matrix.} To alleviate such difficulty, we determine $k$ \textit{candidate skips} from $N$ incoming skips in each searching block by reducing the easy-to-spot ineffective ones. Intuitively, one-to-one relaxation parameters $\alpha$ \cite{liu2018darts} for each skip connection $x$ could be registered and optimized along with the SuperNet. The skip with the highest relaxation score $\alpha$ is then picked as the output of 
$\mathbf{\Phi}(\cdot)$, such that
$\mathbf{\Phi}(\cdot) =x_{\argmax\boldsymbol{\alpha}}$, where 
$\mathbf{x}=\{x_1,\cdots,x_N\}$ and
$\boldsymbol{\alpha}=\{\alpha_1, \cdots, \alpha_N\}$ denote the full set of incoming skips, and their corresponding relaxation scores, respectively. 

However, the above formulation outputs a fixed number of skips rather than flexible ones. We continuously relax skips by constructing a learnable \textit{selection matrix} $\mathbf{M}\in \mathbb{R}^{N \times (N-2)}$ that models the mappings between the $N$ incoming skips and $k$ candidates, and formulate $\mathbf{\Phi}(\cdot)$ as a fully differentiable equation below:
\begin{equation} \label{selectionmatrix}
    \mathbf{\Phi}(\mathbf{x},\mathbf{M})=Matmul(\mathbf{x}, Gumbel\_Softmax(\mathbf{M})),
\end{equation} 

\noindent
where the gumbel-softmax \cite{jang2016categorical} forces each of the $(N-2)$ columns of $\mathbf{M}$ to be an one-hot vector that votes for one of the $N$ incoming skips. Our formulation generates $(N-2)$ selected skips with repetition allowed, achieving a dynamic selection of candidate skips, where the \textit{unique ones} are further averaged out and fed into subsequent blocks. Differing from \cite{liu2018darts,liu2019auto}, we design $\mathbf{\Phi}(\cdot)$ to unify the forward behaviour during both network training and inference stages.

\begin{figure}[t]
	\begin{center}
		\includegraphics[width=1.0\linewidth]{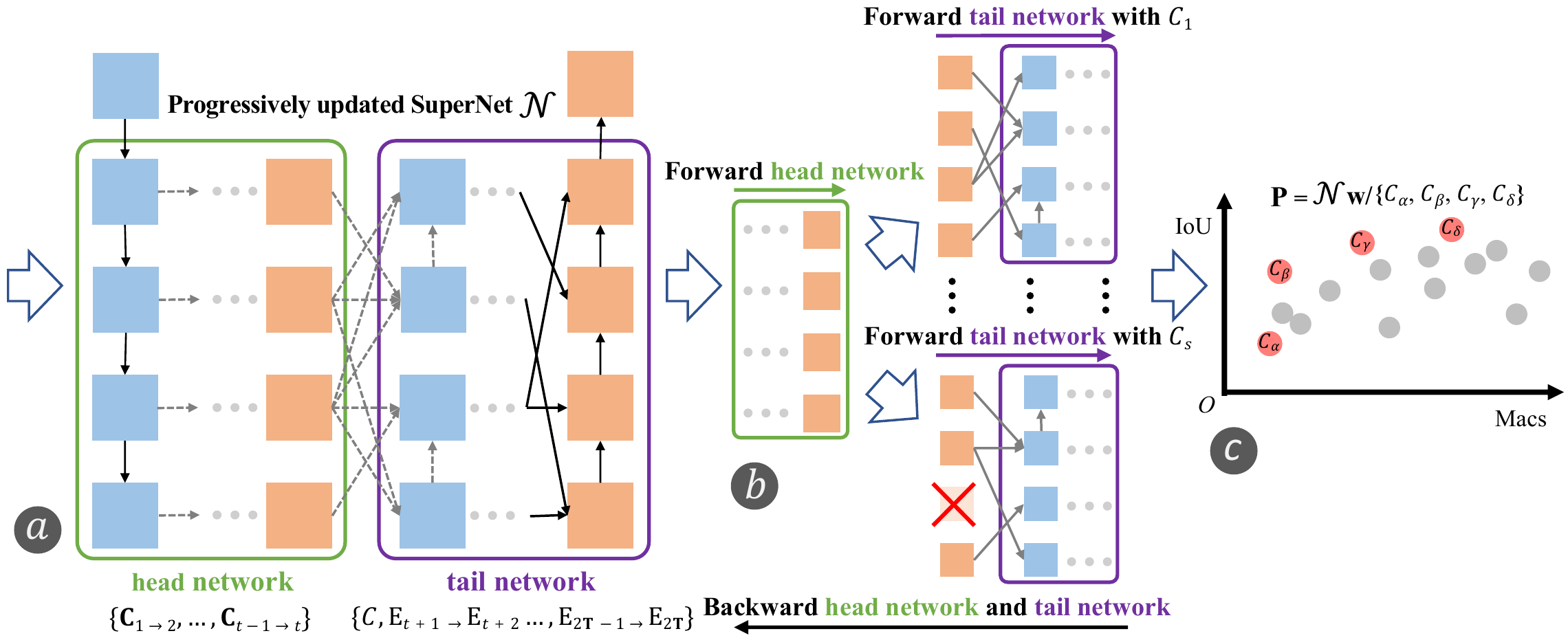}
	\end{center}
	\caption{\textbf{Overview of progressive evolutionary search.} \textbf{(a)} Phase1 searched SuperNet $\mathcal{N}$ can be divided into \textit{head network} and \textit{tail network}. \textbf{(b)} Proposed forward and backward schemes. \textbf{(c)} Only the searched skips at the Pareto front of $\mathbf{P}$ are retained.}
	\label{fig:2}
\end{figure}

\noindent
\textbf{Phase2: Progressive evolutionary search.} 
To further reduce potential redundancies among the candidate skips obtained in Phase1, we perform an additional evolutionary search to find an evolved subset of skips for better network compactness and performance. Specifically, we search the candidate skips for all levels between a certain pair of adjacent extraction stages at the same time, and then progressively move to the next pair once the current search is concluded. As the connectivity of adjacent extraction stages depends on the connectivity of precedent ones, we initiate the search from the last extraction stage pair and progressively move to the first one.

The most straightforward strategies \cite{real2019regularized,real2017large} optimize the SuperNet with each sampling skip set in a population $\mathbf{P}$ individually, and then update $\mathbf{P}$ when all training finish. However, there are two major flaws of such strategies: first, optimizing SuperNet with sampling skips individually may result in unfair outcomes; second, the searching process is empirically slow. Assuming the forward and backward of each extraction stage takes $\mathbf{I_{F}}$ and $\mathbf{I_{B}}$ time, training all $|\mathbf{P}|$ sampling architectures individually for each step takes $2\mathbf{T|P|(I_{F}+I_{B})}$ in total.

\subsection{Analysis of Searching Fairness and Deficiency} \label{analysis}
To overcome the first flaw above, we concept the \textit{skip fairness} and claim that all skip search algorithms need to meet such principle. Note that each sampled architecture $A^i$ (with a subset of skips $C_i$) is randomly drawn from the progressively updated SuperNet $\mathcal{N}$, which makes up the population $\mathbf{P}$ at each iteration.

\begin{algorithm}[t] 
		\caption{Progressive evolutionary search}
		\label{alg1}
		\begin{algorithmic}[1]
			\Require Iteration $\mathbf{T}$, sampling number $\mathbf{s}$, randomly initialized SuperNet weights $\mathcal{W}$, Phase1 searched \textbf{\underline{c}}andidate skips $\{\mathbf{C}_{1\to2}, \cdots, \mathbf{C}_{2\mathbf{T}-1\to2\mathbf{T}}\}$, criterion $\mathcal{L}$. 
			\Ensure BiX-Net with \textbf{\underline{e}}volved skips $\{{\mathbf{E}}_{1\to2}, \cdots, {\mathbf{E}}_{2\mathbf{T}-1\to2\mathbf{T}}\}$
			\For{$t=2\mathbf{T}-1,\cdots,1$}
			\For{$i=1,\cdots,\mathbf{s}$}
			\For{each searching block $b$}
			\State Randomly sample $n$ skips from $\mathbf{C}_{t\to t+1}^b$: $C^b_i, 1\leq n \leq|\mathbf{C}_{t\to t+1}^b|$.
			\EndFor
			\EndFor
			\For{data batch X, target Y}
			\State Forward head network with candidate skips $\mathbf{C}_{1\to2}, \cdots, \mathbf{C}_{t-2\to t-1}$.
			\For{$i=1,\cdots,\mathbf{s}$}
			\State Forward tail network with sampled skips $C_i, {\mathbf{E}}_{t+1\to t+2},\cdots,{\mathbf{E}}_{2\mathbf{T}-1\to 2\mathbf{T}}$.
			\State Calculate loss $l_i=\mathcal{L}$(X, Y)
			\EndFor
			\State Optimize $\mathcal{W}$ with the average loss  $\frac{1}{\mathbf{s}}\sum_{i=1}^{\mathbf{s}}l_i$.
			\EndFor
			\State Get Pareto front from $\{C_1,\cdots,C_\mathbf{s}\}$ and determine ${\mathbf{E}}_{t\to t+1}$.
			\EndFor
		\end{algorithmic}
		
\end{algorithm}

\begin{definition}
\textbf{Skip Fairness.} Let $\mathbf{F} = \{\mathbf{f}^1,\cdots,\mathbf{f}^m\}$ be the skipped features to any searching blocks in each sampled architecture $\mathcal{A}^i$ within a population $\mathbf{P}$. The skip fairness requires $\mathbf{f}^1_{\mathcal{A}^1}\equiv\cdots\equiv\mathbf{f}^1_{\mathcal{A}^{\mathbf{|P|}}},\cdots,\mathbf{f}^m_{\mathcal{A}^1}\equiv\cdots\equiv\mathbf{f}^m_{\mathcal{A}^{\mathbf{|P|}}}$ $\forall \mathbf{f}^i \in \mathbf{F}, \forall \mathcal{A}^i \in \mathbf{P}$.
\end{definition}

The above principle yields that, when searching between the same extraction stage pair, any corresponding level-to-level skips (e.g., $\mathbf{f}^1_{\mathcal{A}^1}$, $\mathbf{f}^1_{\mathcal{A}^2}$, $\mathbf{f}^1_{\mathcal{A}^3}$) across different sampled architectures (e.g., ${\mathcal{A}^1}$, ${\mathcal{A}^2}$, ${\mathcal{A}^3}$) are required to carry identical features. Otherwise, the inconsistent incoming features would impact the search decision on the skipping topology, hence causing unexpected search unfairness. 
Gradient-based search algorithms (e.g. Phase1 search algorithm) meet this principle by its definition, as the same forwarded features are distributed to all candidate skips equally.
However, the aforementioned straightforward strategy violates such principle due to the inconsistent incoming features produced by the individually trained architectures.

Our proposed Phase2 search algorithm meets the skip fairness by synchronizing partial forwarded features in all sampling networks. Specifically, suppose we are searching skips between the $t^{th}$ and $t+1^{th}$ extraction stages $(t\in[1, 2\mathbf{T}-1])$: network topology from the $1^{st}$ to $t-1^{th}$ stages is fixed and the forward process between such stages can be shared. We denote such stages as \textit{head network}. On the contrary, network topology from the $t^{th}$ to $2\mathbf{T}^{th}$ stages varies as the changes of different sampled skips. We then denote such unfixed stages as \textit{tail networks}, which share the same SuperNet weights but with distinct topologies. The forwarded features of head network are fed to all sampling tail networks individually, as shown in Fig. \ref{fig:2}. We average the losses of all tail networks, and backward the gradients through the SuperNet weights only once. Besides, our Phase2 searching process is empirically efficient and overcomes the second flaw above, as one-step training only requires $\mathbf{I_B}+\sum_{t=1}^{2\mathbf{T}-1}(t\mathbf{I_F}+(2\mathbf{T}-t)\mathbf{I_F\cdot|P|})$.

After search between each extraction stage pair completes, we follow a multi-objective selection criterion that retains the architectures at the Pareto front \cite{yang2020cars} based on both validation accuracy (IoU) and computational complexity (MACs). The proposed progressive evolutionary search details are presented in Algorithm \ref{alg1} and the searched BiX-Net is shown in supplementary material Fig. 1.

\section{Experiments}
\subsection{Datasets and Implementation Details}
\noindent
Two segmentation tasks across three different medical image datasets were adopted for validating our proposed method including MoNuSeg \cite{kumar2017dataset}, TNBC \cite{naylor2018segmentation}, and CHAOS \cite{kavur2021chaos}. \textbf{(1)} The MoNuSeg dataset contains 30 training images and 14 testing images of size $1000\times1000$ cropped from whole slide images of different organs. Following \cite{xiang2020bio}, we extracted $512\times512$ patches from the corners of each image. \textbf{(2)} The TNBC dataset was used as an \textit{extra validation dataset} \cite{xiang2020bio,graham2018hover}, consisting of $512\times512$ sub-images obtained from 50 histopathology images of different tissue regions. \textbf{(3)} We also conducted 5-fold cross-validation on MRI scans from the CHAOS dataset to evaluate the generalization ability of BiX-Net on the abdominal organ segmentation task, which contains 120 DICOM image sequences from T1-DUAL (both in-phase and out-phase) and T2-SPIR with a spatial resolution of $256 \times 256$.
We pre-processed the raw sequences by min-max normalization and auto-contrast enhancement.
\\
\noindent
\textbf{Searching Implementation.} 
We utilized MoNuSeg only for searching the optimal BiX-Net, and the same architecture is then transferred to all other tasks. We followed the same data augmentation strategies as in \cite{xiang2020bio}. In Phase1, the BiO-Net++ SuperNet was trained with a total of 300 epochs with a base learning rate of 0.001 and a decay rate of $3e^{-3}$. The network weights and the selection matrices were optimized altogether with the same optimizer, rather than optimized separately \cite{liu2018darts}. Our Phase1 searching procedure took only 0.09 GPU-Day. In Phase2, there were total 5 searching iterations when $\mathbf{T}=3$. At each iteration, for each retained architecture from the preceding generation, we sampled $s=15$ different skip sets to form the new population $\mathbf{P}$ that were trained 40 epochs starting from a learning rate of $0.001$ and then decayed by 10 times every 10 epochs. Due to GPU memory limitation, we only retained $|\mathbf{P}| < 3$ architectures with highest IoU at the Pareto front. Our Phase2 searching process consumed 0.37 GPU-Day. \\
\textbf{Retraining Implementation.} Adam optimizer \cite{kingma:adam} was used across all experiments to minimize cross entropy loss in both network searching and retraining. Batch size was set to be 2 for MoNuSeg and TNBC, and 16 for CHAOS. 
We retrained the constructed BiX-Net and all competing models with the same implementation as Phase1, which was identical across all experiments for fair comparisons. Mean intersection of Union (mIoU) and Dice Coefficient (DICE) were reported to evaluate accuracy while Multiplier Accumulator (MACs) was reported to measure computational complexity. 

\begin{table}[t]
		\caption{Comparison on MoNuSeg and TNBC with three independent runs.}\label{nuclei}
		\centering
		\begin{tabular}{{|l|c c|c c|c | c|}}
			\toprule
			\multicolumn{1}{c}{} & \multicolumn{2}{c}{MoNuSeg} & \multicolumn{2}{c}{TNBC}& 	\multicolumn{2}{c}{}\\
			\hline
			Methods &  mIoU(\%) & DICE(\%) & mIoU(\%) & DICE(\%) &Params(M) &MACs(G)\\
			\hline
			\hline
			U-Net \cite{ronneberger2015u} & 68.2$\pm$0.3 & 80.7$\pm$0.3 & 46.7$\pm$0.6 & 62.3$\pm$0.6 & 8.64 & 65.83\\
			R2U-Net \cite{alom2018nuclei} & 69.1$\pm$0.3 & 81.2$\pm$0.3 & 60.1$\pm$0.5 & 71.3$\pm$0.6 & 9.78 & 197.16\\
			BiO-Net \cite{xiang2020bio} & 69.9$\pm$0.2 & 82.0$\pm$0.2 & 62.2$\pm$0.4 & 75.8$\pm$0.5 &14.99& 115.67\\
			\hline
			NAS-UNet \cite{weng2019unet}&68.4$\pm$0.3&80.7$\pm$0.3&54.5$\pm$0.6&69.6$\pm$0.5&2.42&67.31\\
			AutoDeepLab \cite{liu2019auto} &68.5$\pm$0.2&81.0$\pm$0.3& 57.2$\pm$0.5 & 70.8$\pm$0.5 &27.13&60.33\\
			MS-NAS \cite{yan2020ms} &68.8$\pm$0.4&80.9$\pm$0.3&58.8$\pm$0.6&71.1$\pm$0.5&14.08&72.71\\
			\hline
			BiO-Net++ &70.0$\pm$0.3&82.2$\pm$0.3&67.5$\pm$0.4&80.4$\pm$0.5&0.43&34.36\\
			Phase1 searched &69.8$\pm$0.2&82.1$\pm$0.2&66.8$\pm$0.6&80.1$\pm$0.4&0.43&31.41\\
			\hline
			\textbf{BiX-Net} & \textbf{69.9$\pm$0.3} & \textbf{82.2$\pm$0.2} & \textbf{68.0$\pm$0.4} & \textbf{80.8$\pm$0.3} & \textbf{0.38} & \textbf{28.00}\\
            \hline
		\end{tabular}
	\end{table}
	
	\begin{table}[t]
		\caption{Comparison on CHAOS (MRI) with 5-fold cross validation.}\label{chaos}
		\centering
		\resizebox{\linewidth}{!}{\begin{tabular}{{|l|c c|c c|c c|c c|}}
			\toprule
			\multicolumn{1}{c}{} & \multicolumn{2}{c}{Liver} & \multicolumn{2}{c}{Left Kidney}& 	\multicolumn{2}{c}{Right Kidney}&\multicolumn{2}{c}{Spleen}\\
			\hline
			Methods &  mIoU(\%) & DICE(\%) & mIoU(\%) & DICE(\%) &mIoU(\%)&DICE(\%) & mIoU(\%) & DICE(\%)\\
			\hline
			\hline
			U-Net  &78.1$\pm$2.0& 86.8$\pm$1.8& 61.3$\pm$1.1 &73.8$\pm$1.2 &63.5$\pm$1.1 &76.2$\pm$1.1 &62.2$\pm$2.1&74.4$\pm$2.3\\
			BiO-Net  &  85.8$\pm$2.0& 91.7$\pm$1.8 & 75.7$\pm$1.1& 85.1$\pm$1.2&78.2$\pm$1.0&87.2$\pm$1.1&73.2$\pm$2.3&82.8$\pm$2.3\\
			\hline
			NAS-UNet  & 79.1$\pm$1.8&87.2$\pm$1.8&65.5$\pm$1.5&75.0$\pm$1.3&66.2$\pm$1.2&77.7$\pm$1.0&64.1$\pm$1.3&75.8$\pm$1.6\\
			AutoDeepLab  & 79.8$\pm$1.9&88.1$\pm$1.8&66.7$\pm$1.6&75.0$\pm$1.7&61.9$\pm$0.9&75.7$\pm$1.1&63.9$\pm$1.2&75.5$\pm$1.4\\
			MS-NAS & 72.6$\pm$2.3&82.6$\pm$2.1&71.0$\pm$1.3&81.9$\pm$1.3&70.1$\pm$1.9&81.1$\pm$1.8&62.5$\pm$2.1&74.0$\pm$2.3\\
			\hline
			\textbf{BiX-Net} & \textbf{82.6$\pm$1.5}&\textbf{89.8$\pm$1.5}&\textbf{71.0$\pm$1.0}&\textbf{82.1$\pm$1.1}&\textbf{71.9$\pm$0.8}&\textbf{82.7$\pm$1.0}&\textbf{66.0$\pm$1.7}&\textbf{76.5$\pm$2.0}\\
			\hline
		\end{tabular}}
	\end{table}
	
\begin{figure}[t]
	\begin{center}
		\includegraphics[width=1.0\linewidth]{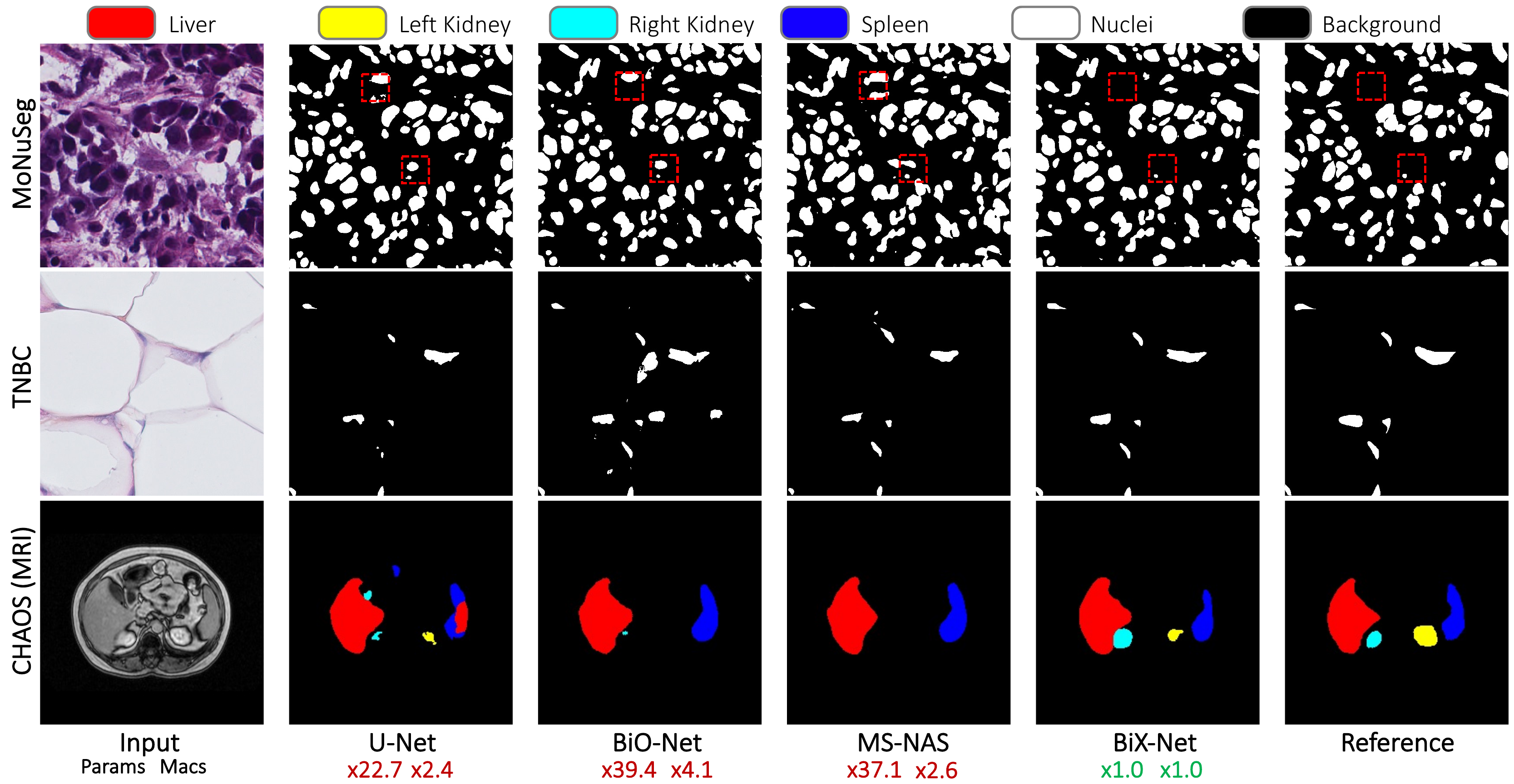}
	\end{center}
	\caption{\textbf{Qualitative comparison of different segmentation results.} Relative parameter (left) and computation overheads (right) compared to BiX-Net are also shown.}
	\label{fig:vis1}
\end{figure}

\subsection{Experimental Results}
\textbf{Nuclei Segmentation.} 
Our method was compared to the vanilla U-Net \cite{ronneberger2015u}, state-of-the-art recurrent U-Net variants \cite{alom2018nuclei,xiang2020bio}, and homogeneous state-of-the-art NAS searched networks \cite{weng2019unet,liu2019auto,yan2020ms}.
All models were trained from scratch with final results reported as the average of three independent runs. Table \ref{nuclei} shows that our plain BiX-Net outperforms the state-of-the-art NAS counterparts considerably, and achieves on par results to BiO-Net \cite{xiang2020bio} with significantly lower network complexity. Noteworthy, our results are much higher than all others on TNBC, which indicates superior generalization ability. Qualitative comparisons on all datasets are shown in Fig. \ref{fig:vis1}. 

\noindent
\textbf{Multi-class Organ Segmentation.}
CHAOS challenge aims at the precise segmentation of four abdominal organs separately: liver, left kidney, right kidney, and spleen in a CT or MRI slice. 
Instead of training networks on each class as several independent binary segmentation tasks \cite{yan2020ms}, we reproduced all models to output the logits for all classes directly. Similar to the nuclei segmentation, Table \ref{chaos} indicates that BiX-Net achieves the best performance of all classes among state-of-the-art NAS searched networks \cite{weng2019unet,liu2019auto,yan2020ms} with much lower computational complexity. Although the hand-crafted BiO-Net outperforms all comparison methods, it suffers from the computation burdens, which are a 4.1 times of computational complexity, and a 39.4 times of trainable parameter number. 
Additionally, BiX-Net produces much better segmentation mask when all organs are presented in a single slice (Fig. \ref{fig:vis1}).

\noindent
\textbf{Ablation Studies.}
In addition, we conducted two ablation studies by training the presented SuperNet BiO-Net++ (the multi-scale upgrade of BiO-Net) and the Phase1 searched architecture (has not been searched by Phase2) directly on nuclei datasets. As shown in Table \ref{nuclei}, our Phase1 search algorithm provides a 8.6\% MACs reduction compared to BiO-Net++, and our BiX-Net eventually achieves a 18.5\% MACs reduction, which validates the necessity of Phase1 and Phase2 search of BiX-NAS. Unlike prior NAS works, our finally searched BiX-Net follows the recurrent bi-directional paradigm with repeated use of the same building blocks at different iterations. Note that there is one building block (encoder at level 4) that has been skipped at all iterations (supplementary material Fig. 1.), resulting in a further reduction in the total network parameters. 

For all metrics, BiX-Net obtains higher scores than competing NAS counterparts on all datasets and achieves on-par results with our proposed BiO-Net++ with fewer computations. Additionally, we perform two tail paired t-test to analyze the statistical significance between our method and other competing NAS methods. BiX-Net achieves p-values $<0.05$ on the nuclei datasets and $<0.1$ on the CHAOS dataset, validating the significance of our method.

\section{Conclusion}
In this work, we proposed an efficient two-phase NAS algorithm that searches for bi-directional multi-scale skip connections between encoder and decoder, namely BiX-NAS. We first follow differentiable NAS with a novel selection matrix to narrow down the search space. An efficient progressive evolutionary search is then proposed to further reduce skip redundancies. Experimental results on various segmentation tasks show that the searched BiX-Net surpasses state-of-the-art NAS counterparts with considerably fewer parameters and computational costs.

%
%
%
\bibliographystyle{splncs04}
\bibliography{ref}
%




\end{document}